\documentclass[12pt,a4paper]{article}
\usepackage{amsfonts,latexsym}
\usepackage{amsmath,amssymb}
\usepackage{graphicx,color}

\renewcommand{\thefootnote}{\fnsymbol{footnote}}

\begin{document}

\vspace{12mm}

\begin{center}
{{{\Large {\bf Thermodynamic and shadow radius  analysis  of the charged Einstein-Euler-Heisenberg black hole }}}}\\[10mm]

{Yun Soo Myung\footnote{e-mail address: ysmyung@inje.ac.kr}}\\[8mm]

{Center for Quantum Spacetime, Sogang University, Seoul 04107, Republic of  Korea\\[0pt] }

\end{center}
\vspace{2mm}

\begin{abstract}
We perform the thermodynamic and  shadow radius analysis of an electrically charged  black hole (EC) with electric charge $q$ and coupling constant $\mu$ obtained from the Einstein-Euler-Heisenberg nonlinear electodynamics.
For $\mu=0.03$,  we have four solution branches of the horizon including low, hot, negative, and cold ones, while for $\mu=0.3,3$ there exist single branches without limitation on $q$.
The shadow radius for the low branch is the nearly same as that for the Reissner-Norstr\"om black hole for $q<1$ case,  while one finds the $q>1$ negative branch which is constrained by the EHT observation.
\end{abstract}
\vspace{5mm}

\vspace{1.5cm}

\hspace{11.5cm}{Typeset Using \LaTeX}
\newpage
\renewcommand{\thefootnote}{\arabic{footnote}}
\setcounter{footnote}{0}


\section{Introduction}

Supermassive black holes founded at the center of galaxies have played the important role in galaxy formation and galaxy evolution.
The images of the M87* BH~\cite{EventHorizonTelescope:2019dse,EventHorizonTelescope:2019ths,EventHorizonTelescope:2019ggy} have inspired enormous  studies on the  BH.
The recent EHT observation has focussed on  the center of our galaxy and shown  promising  images of the  SgrA* BH~\cite{EventHorizonTelescope:2022wkp,EventHorizonTelescope:2022wok,EventHorizonTelescope:2022xqj}.
The BH images  indicated that there is a dark central region  surrounded by a bright ring, which are called shadow and photon sphere (light ring) of the BH, respectively.
Strong deflection lensing can generate a shadow and relativistic images caused by photons winding several loops around a BH~\cite{Bardeen:1972fi,Bozza:2010xqn}. The photon sphere of a BH  plays an important role in the strong deflection.
Interestingly, the shadow of BH with scalar hair was employed  to test the EHT results~\cite{Khodadi:2020jij}, while the shadows of other BHs, wormholes, and naked singularities  obtained  from various modified gravity theories have been used  to constrain their hair parameters~\cite{Vagnozzi:2022moj}.

It is well known that the BH thermodynamics  is universal and can be  applicable to BHs found in various modified gravity theories. If one knows the entropy and temperature for a BH, one finds its thermal stability through the heat capacity.
Also, if one has a few  solution branches to the horizon, computation of their entropy and temperature makes the nature of branches clear~\cite{Blazquez-Salcedo:2020nhs,LuisBlazquez-Salcedo:2020rqp}.

In this work, we wish to  perform thermodynamic and  shadow radius analysis of an electrically charged black hole (EC) with electric charge $q$ and coupling constant $\mu$  obtained from the Einstein-Euler-Heisenberg-nonlinear electrodynamics (EEH)~\cite{Yajima:2000kw,Amaro:2020xro,Breton:2021mju,Allahyari:2019jqz}.
Recently, there were thermodynamic  studies~\cite{Magos:2020ykt,Zhao:2024phz,Gursel:2025wan}, weak deflection angle~\cite{Fu:2021akc}, and optical appearance~\cite{Zeng:2022pvb}  on the black hole with electric (or magnetic) charges.

 Our analysis depends critically  on the coupling constant $\mu$.
 For $\mu=0.03$,  we have four solution branches of the horizon including low, hot, negative, and cold ones, while for $\mu=0.3,3$ there exist single branches.
 There are naked singularity (NS) arisen  from the charge extension of the  light rings for the negative and cold branches.
 It is important to note that  for $\mu=0.3,3$, there is no limitation on the magnetic charge $q$ for its horizon, thermodynamic quantities, light ring, and shadow radius.
  Therefore, there is no NS arisen  from the charge extension of the  light rings for $\mu=0.3,3$.   The shadow radii for the EC are the nearly same as that for the Reissner-Norstr\"om  black hole (RN) for $q<1$, while one finds the $q>1$ negative branch  which is constrained by the recent EHT observation. The analysis of shadow radius   was discussed in the magnetically charged black hole~\cite{Allahyari:2019jqz,Vagnozzi:2022moj}.

\section{Electrically charged  black holes}

We introduce  the  Einstein-Euler-Heisenberg nonlinear electodynamics (EEH) with $G=1$~\cite{Magos:2020ykt}
\begin{equation}
{\cal L}_{\rm EEH}=\frac{1}{16\pi}\Big[R-4F+2a\Big(F^2+\frac{7}{4}G^2\Big)\Big],\quad F=\frac{F_{\mu\nu}F^{\mu\nu}}{4},\quad G=\frac{{}^*F^{\mu\nu}F_{\mu\nu}}{4}
\end{equation}
where a coupling constant $a=\frac{8\alpha^2}{45m_e^4}$ is redefined as $8\mu$ with $\mu=\frac{\alpha^2}{45m_e^4}$ to recover the same metric function for a magnetically charged black hole (MC)~\cite{Allahyari:2019jqz}.

To find the black hole solution, we consider  a spherically symmetric  spacetime and tensor $P_{\mu\nu}$ as
\begin{equation}\label{metric-ansatz}
ds^2_{\rm EEH}=-f(r)dt^2+\frac{dr^2}{f(r)}+r^2(d\theta^2+\sin^2\theta d\phi^2),\quad P_{\mu\nu}=\frac{q^2}{r^2}\delta^0_{[\mu}\delta^1_{\nu]}
\end{equation}
with $q$ an electric charge.
Here,  the metric function  takes the forms
\begin{equation} \label{m-func}
f(r)=1-\frac{2m}{r}+\frac{q^2}{r^2}-\frac{2\mu}{5} \frac{q^4}{r^6}
\end{equation}
which recovers the RN in the limit of $\mu\to 0$.

From $f(r)=0$ with $\mu=0.03$, one finds that  there are four  branches (positive roots) of the  horizon
\begin{equation}
r_{L}(m,q,\mu),\quad r_{H}(m,q,\mu),\quad r_{N}(m,q,\mu),\quad r_{C}(m,q,\mu)
\end{equation}
whose forms are too complicated to show here. Here, $q-(q+)$ represent $q\in[0,0.95](q\in[0.95,$ ]).  For $\mu=0.3,3$, there exist  single branches of $r_L(m,q,\mu)$. The names of subscript are clear when its thermodynamic quantities are found: L, H, N, C imply low, hot, negative, cold, respectively. The case of $\mu=0.03$ is similar to Einstein-Maxwell-scalar black holes with a quartic coupling function $f(\Phi)=1+\alpha \Phi^4$:  the hot, the cold, and the bald (RN) branches ~\cite{Blazquez-Salcedo:2020nhs,LuisBlazquez-Salcedo:2020rqp}.
\begin{figure*}[t!]
   \centering
   \includegraphics[width=0.4\textwidth]{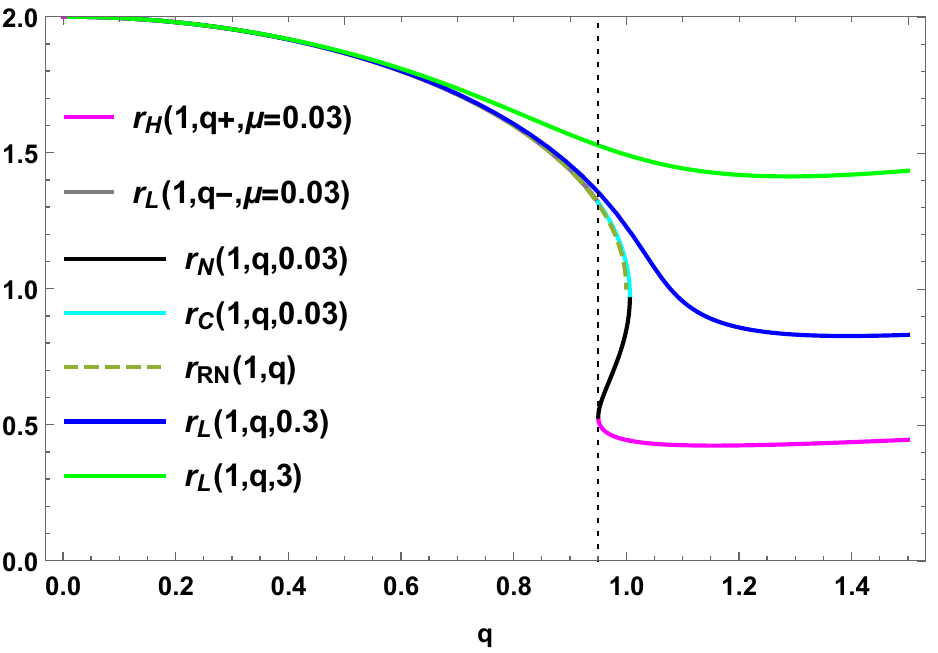}
   \hfill%
    \includegraphics[width=0.4\textwidth]{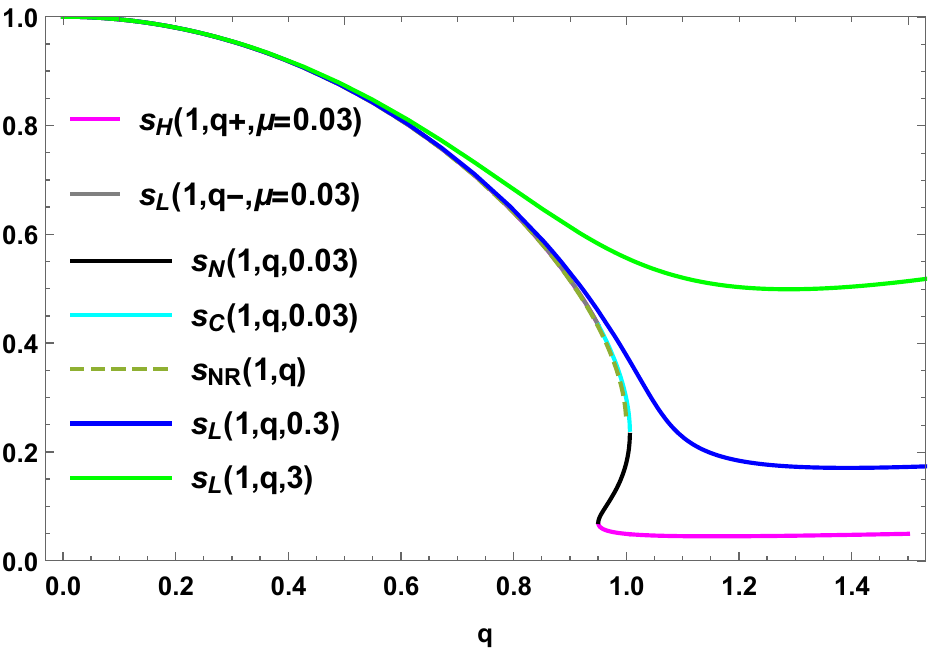}
\caption{ (Left) Four branches of the horizon  $r_{L}(1,q-,\mu=0.03)\ge r_{C}(1,q,0.03)\ge r_{N}(1,q,0.03)\ge r_{H}(1,q+,0.03)$  as functions of $q$ with $r_{RN}(1,q)$. Here, we  note that $r_{L}(1,q,0.3)$ and  $r_{L}(1,q,3)$ are  defined as  single horizons without limitation of $q$. A dotted line represents $q=0.95$ where the low branch  meets the cold one  and the negative branch  meets the hot one.   (Right) Four reduced entropies $s_{L}(1,q-,\mu=0.03)\ge s_{C}(1,q,0.03)\ge s_{N}(1,q,0.03)\ge s_{H}(1,q+,0.03)$  as functions of $q$ with $s_{RN}(1,q)$.   }
\end{figure*}
For a reference, we wish to include the RN solution with its event horizon and electric potential
\begin{equation}
r_{RN}(m,q)=m-\sqrt{m^2-q^2},\quad A=\frac{qdt}{r},
\end{equation}
which implies  $0<q/m < 1$ for existing two (event/Cahuchy) horizons.
From (Left) Fig. 1, one finds that $r_{L}(1,q,0.3)$ and $r_{L}(1,q,3)$  have  single horizons without limitation of $q$.   An important  point  is
that there is no  theoretical constraint on restricting  the electric charge $q$.
On the other hand, one may have four branches for  $\mu=0.03$ which depend on the $q$ range: $r_L(1,q-,0.03),~r_C(1,q\in[0.95,1.0065],0.03),~r_N(1,q\in[1.0065,0.95],0.03),~r_{H}(1,q+,0.03)$.
 A dotted line represents $q=0.9503(\simeq 0.95)$ where the low branch  meets the cold one  and the negative branch  meets the high one. At $q=1.0065$, the cold branch meets the negative one.
\begin{table}[h]
\begin{tabular}{|c|c|c|c|c|c|c|c|}
  \hline
  $q$& 0.5  & 0.6  & 0.7&0.85&0.9&0.95 \\ \hline
  $\mu$ & (0,0.0002]&(0,0.001]&(0,0.002] &(0,0.011]&(0,0.018]&(0, 0.03] \\ \hline
   $q$&1&1.0065&1.01&1.015& 1.02&1.021 \\ \hline
  $\mu$&(0,0.055] &[0.03,0.06]&[0.044,0.064]&[0.062,0.069]&[0.0758,0.076]& N.A. \\ \hline
\end{tabular}
\caption{$\mu$-region for the existence of $r_N(1,q,\mu)$ and $r_C(1,q,\mu)$, depending on the electric charge $q$.  }
\end{table}

It is important to clarify the $\mu$ and $q$-regions  for the existence of $r_N(m,q,\mu)$ and $r_C(m,q,\mu)$ with $m=1$. We compute them numerically because there is no analytic result.
In Table 1, we display the coupling parameter $\mu$-region for the existence of  $r_N(1,q,\mu)$ and $r_N(1,q,\mu)$,  depending on the electric charge $q$. For $q=0.1$, its allowed $\mu$-range is tiny like as $(0,1.23\times 10^{-8}]$.   As $q$ increases from $q=0.5$, the $\mu$-range increases. As $q$ increases from $q=1$, the $\mu$-range becomes smaller and smaller. Its $q$-upper limit is $q=1.02$ and the corresponding  $\mu$-region is small like as [0.0758,0.076] which is less than 0.1 for a given $m=1$. This is why we choose $\mu=0.03$ for realizing four solution  branches of the horizon.  It is worth noting that $\mu$  does not exist beyond  $q=1.02$.  For $q\in[0.95,1.0065]$, $\mu=0.03$ corresponds to  the upper bound as well as the lower bound, guaranteeing the existence of $r_N(1,q,\mu)$ and $r_C(1,q,\mu)$ as shown in (Left) Fig. 1. If one chooses a smaller $\mu<0.03$, two branches of $r_N(1,q,\mu)$ and $r_C(1,q,\mu)$ exist for a smaller $q$-range than $q\in[0.95,1.0065]$.
This shows clearly how to develop four solution branches of the horizon  according to the $\mu$-term ($-2\mu q^4/5r^6$) in the metric function Eq.(\ref{m-func}).

\section{Thermodynamic analysis}
The mass is obtained from $f(r_{i})=0$ for $i$=L,H,N,C as
\begin{eqnarray}
M_i(m,q,\mu)&=&\frac{r_{i}(m,q,\mu)}{2}\Big[1+ \frac{q^2}{r_{i}^2(m,q,\mu)}-\frac{2\mu q^4}{5r_{i}^6(m,q,\mu)}\Big]. \label{mass-1}
\end{eqnarray}
Also, the temperature from the surface gravity and chemical (electric  and $\mu$) potentials at the horizon  from $\partial M_i(m,q,\mu)/\partial q$ and $\partial M_i(m,q,\mu)/\partial \mu$ are given by
\begin{eqnarray}
T_{i}(m,q,\mu)&=&\frac{\kappa}{2\pi}=\frac{f'(r_i)}{4\pi}=\frac{1}{4\pi r_{i}(m,q,\mu)}\Big[1- \frac{q^2}{r_{i}^2(m,q,\mu)}-\frac{2\mu q^4}{r_{i}^6(m,q,\mu)}\Big], \label{sg1}\\
W_{q,i}(m,q,\mu)&=&\frac{q}{r_{i}(m,q,\mu)}-\frac{4\mu q^3}{5r^5_{i}(m,q,\mu)},\quad W_{\mu,i}(m,q,\mu)=-\frac{q^4}{5r_i^5(m,q,\mu)}. \label{sg2}
\end{eqnarray}
Here, we note that $T_{i}(m,q,\mu)$ can be  obtained from replacing $S_i$ by $ \pi r_i^2$ after  considering $\partial M_i(S_i,q,\mu)/\partial S_i$.
With the area-law entropy   and temperature $T_{i}$, we check  that the first law~\cite{Gursel:2025wan}
\begin{equation}
dM_i(m,q,\mu)=T_{i}(m,q,\mu) dS_i(m,q,\mu)+W_{q,i}(m,q,\mu)dq+W_{\mu,i}(m,q,\mu)d\mu \label{1st-law}
\end{equation}
is satisfied. Furthermore, it is shown that the Smarr formula expressing the relation between the thermodynamic quantities
 is satisfied  as
\begin{equation}
M_i(m,q,\mu)=2T_{i}(m,q,\mu)S_i(m,q,\mu)+W_{q,i}(m,q,\mu)q+2W_{\mu,i}(m,q,\mu)\mu.
\end{equation}
In (Right) Fig. 1, we display the reduced entropy defined by $s_i(m,q,\mu)=S_{i}(m,q,\mu)/4\pi m^2$ which shows a similar behavior as in (Left) Fig. 1. The reduced temperature defined by $t_i(m,q,\mu)=8\pi m T_{i}(m,p,q)$ is shown in Fig. 2. This shows clearly why we use the subscripts  $L,H,C,N$, representing  low, high, cold, negative temperatures, respectively.
A dotted line denotes $q=0.95$ at which  five temperatures pass it.
\begin{figure*}[t!]
   \centering
  \includegraphics[width=0.4\textwidth]{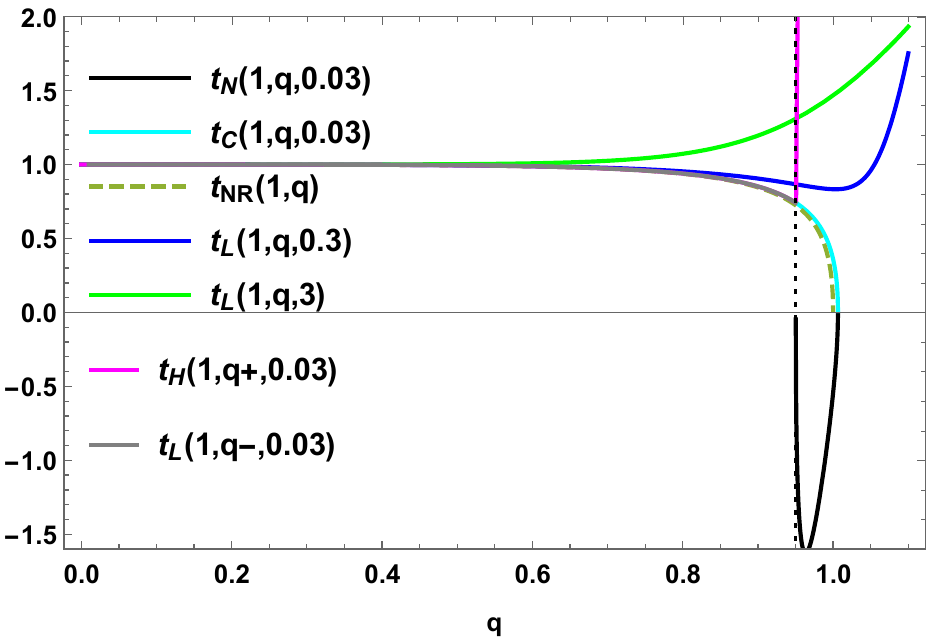}
 \hfill%
    \includegraphics[width=0.4\textwidth]{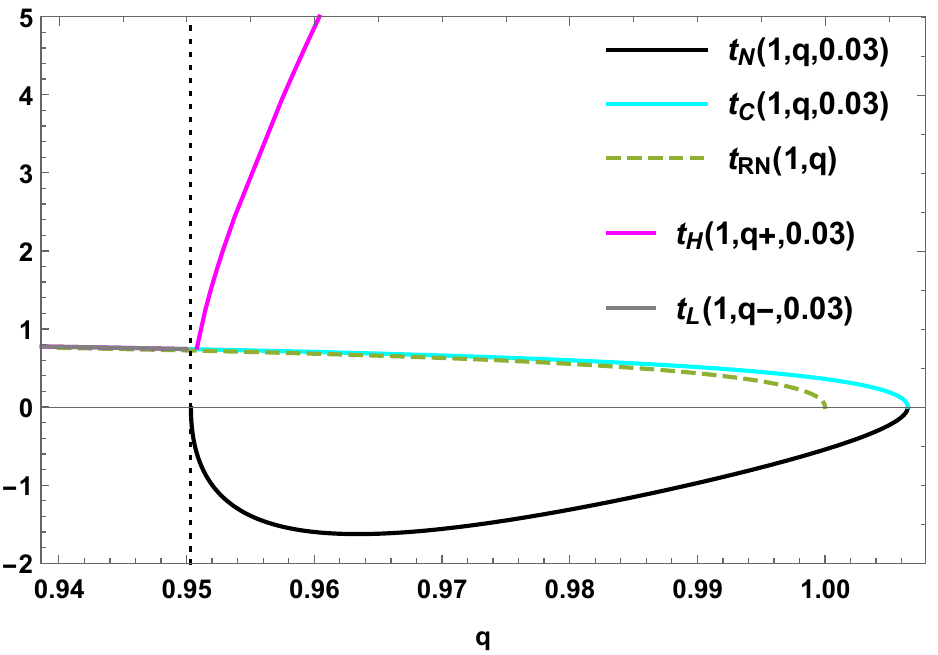}
\caption{(Left) Several  reduced temperatures $t_i(m=1,q,\mu=0.03)$ for $i$=L,N,C,H, $t_{L}(1,q,0.3)$, $t_{L}(1,q,3)$, and $t_{RN}(1,q)$ as functions of $q\in[0,1.2]$.
(Right) Enlarged reduced temperatures $t_i(1,q,0.03)$ and $t_{RN}(1,q)$ are  functions of  $q\in[0.94,1.0065]$. This shows clearly that L,H,C,N represent low, hot, cold, negative temperatures, respectively.
A dotted line denotes $q=0.95$ where five temperatures pass it. }
\end{figure*}

Now, we are in a position to introduce  the heat capacity defined by $\partial M_i/\partial t_i$ as
\begin{equation}
c_i(m,q,\mu)=-\frac{r_i^2(m,q,\mu)[2\mu q^4-q^2 r_i^4(m,q,\mu)+r_i^6(m,q,\mu)]}{4[14\mu q^4-3q^2r_i^4(m,q,\mu)+r_i^6(m,q,\mu)]}.
\end{equation}
We note that the locally thermal  stability (instability) can be achieved when $c_i>0(c_i<0)$.
\begin{figure*}[t!]
   \centering
  \includegraphics[width=0.4\textwidth]{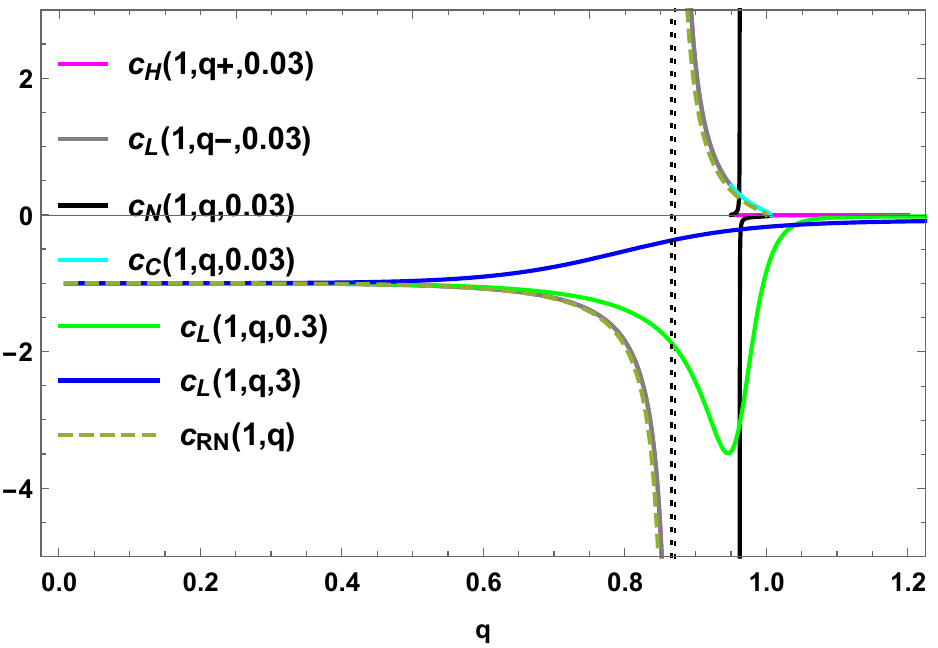}
 \hfill%
    \includegraphics[width=0.4\textwidth]{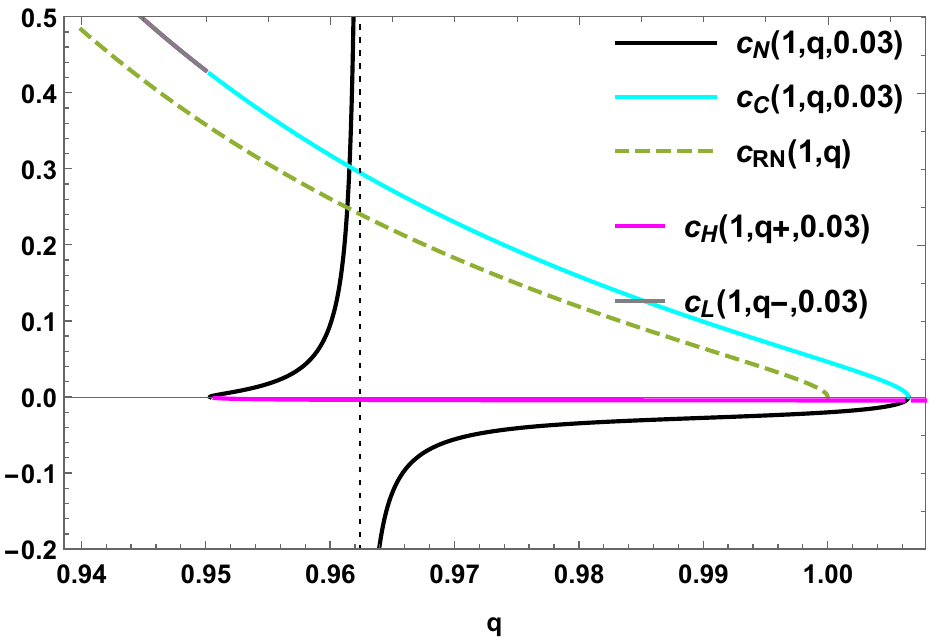}
\caption{(Left) Several  heat capacities  $c_i(m=1,q,\mu=0.03)$ for $i$=L,N,C,H, $c_{L}(1,q,0.3)$, $c_{L}(1,q,3)$, and $c_{RN}(1,q)$ as functions of $q\in[0,1.2]$.  There are two Davies points  (dotted lines at $q=0.866,0871$). and two extremal points with $c_{i}=0$.  (Right) Enlarged heat capacities  $c_i(1,q,0.03)$ and $c_{RN}(1,q)$ are  functions of  $q\in[0.94,1.0065]$. The dotted line indicates a Davies line at $q=0.962$  for the negative branch and  three extremal points appear as $c_{i}(1,q,0.03)=0$ with $c_{RN}(1,q)=0$. }
\end{figure*}
In addition, the reduced RN temperature and heat capacity are given by
\begin{eqnarray}
t_{RN}(m,q)&=&\frac{2}{ r_{ RN}(m,q)}\Big[1-\frac{q^2}{r_{RN}^2(m,q)}\Big], \label{RN-t} \\
c_{RN}(m,q)&=&-\frac{r_{RN}^2(m,q)}{4}\frac{(r_{RN}^2(m,q)-q^2)}{(r_{ RN}^2(m,q)-3q^2)}.\label{RN-hc}
\end{eqnarray}
From Fig. 3, one finds that three Davies lines for $c_{RN}(1, q),
   c_L(1, q-, 0.03), c_N(1, q, 0.03)$  are located  at $q = 0.866, 0.871, 0.962$,
$c_C \ge  0$, and  $c_{H} (1, q +, 0.03) \le  0,~c_L (1, q, 0.3)\le  0,~c_L (1, q, 3) \le  0$. This means that three branches including Davies points  make phase transitions, the cold branch is thermodynamically stable,
whereas the last  three branches are always  thermodynamically unstable. We have thermodynamically stable regions: $q\in[0.871,0.95]$ for the low branch, $q\in[0.95,1.0065]$ for the cold branch, $q\in[0.95,0.962]$ for the negative branch, and $q\in[0.866,1]$ for the RN branch. These regions might be related to the shadow radii predicted by the EHT observation.
\begin{figure*}[t!]
   \centering
  \includegraphics[width=0.4\textwidth]{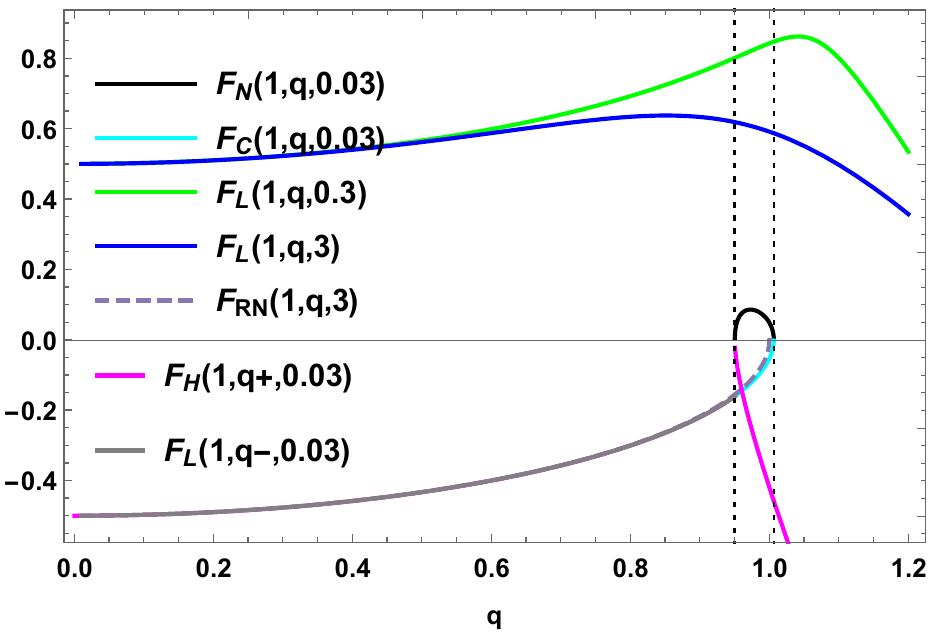}
\caption{ Several Helmholtz free energies   $F_i(m=1,q,\mu=0.03)$ for $i$=L,N,C,H, $F_{L}(1,q,0.3)$, $F_{L}(1,q,3)$, and $c_{RN}(1,q)$ as functions of $q\in[0,1.2]$.   Two dotted lines indicate the zero free energy at $q=0.95$ and 1.0065 where the latter is considered as an extremal point.  }
\end{figure*}
Finally, we need  the Helmholtz free energy to test the global stability. If it is positive (negative), it is globally unstable (stable). In this case, we introduce the ground state as an extremal black hole for $\mu=0.03$ like as the RN case.  The Helmholtz free energy is defined as $F_i=M_i-T_iS_i-M_{e,i}$ with the RN free energy
\begin{eqnarray}
F_i(m,q,\mu)&=&\frac{r_i(m,q,\mu)}{4}+ \frac{3q^2}{r_{i}(m,q,\mu)}-\frac{7\mu q^4}{10r_{i}^5(m,q,\mu)}-M_{e,i}. \label{f1-energy} \\
F_{RN}(m,q)&=&\frac{r_{RN}(m,q)}{4}+ \frac{3q^2}{r_{RN}(m,q)}-M_{e,RN}. \label{frn-energy}
\end{eqnarray}
As is shown in Fig. 4, the Helmholtz free energy for $\mu=0.3,3$ are positive for $q\in[0,1.2]$ because they  do not include extremal points. This means that they are globally unstable.
However, three branches of $r_{L}(1,q-,0.03),r_{H}(1,q+,0.03),r_{C}(1,q,0.03)$ with $r_{RM}(1,q)$ are globally stable except that  the negative branch is globally unstable.

\section{Light rings and shadow radii}
We introduce the Lagrangian of the photon to find the light rings of the EC
\begin{equation}
{\cal L}_{\rm LR}=\frac{1}{2}g_{\mu\nu}\dot{x}^\mu\dot{x}^\nu=\frac{1}{2}\Big[-f(r)\dot{t}^2+\frac{\dot{r}^2}{f(r)}+r^2(\dot{\theta}^2+\sin^2\theta \dot{\phi}^2)\Big].
\end{equation}
Taking the light traveling on the equational plane of the  EC ($\theta=\pi/2$ and $\dot{\theta}=0$) described by  a spherically symmetric and static metric Eq.(\ref{metric-ansatz}),
two conserved quantities of photon (energy and angular momentum) are given by
\begin{equation}
E=-\frac{\partial {\cal L}_{LR}}{\partial \dot{t}}=g(r)\dot{t},\quad \tilde{\ell}=\frac{\partial {\cal L}_{LR}}{\partial \dot{\phi}}=r^2\dot{\phi}.
\end{equation}
Choosing the null geodesic for the photon ($ds^2=0$) with the affine parameter $\tilde{\lambda}=\lambda \tilde{\ell}$ and impact parameter $\tilde{b}=\tilde{\ell}/E$, its radial equation of motion is  given by
\begin{equation}
\frac{dr}{d\tilde{\lambda}}=\sqrt{\frac{1}{\tilde{b}^2}-\frac{f(r)}{r^2}}.
\end{equation}
Here, the effective potential for a photon takes the form
\begin{equation}
V(r)=\frac{f(r)}{r^2}.
\end{equation}
Requiring the light ring (photon sphere: $\dot{r}=0,~\ddot{r}=0$), one finds two conditions
\begin{equation} \label{cond-LR}
V(r=L)=\frac{1}{2b^2}, \quad V'(r=L)=0,
\end{equation}
where $b$ denotes the critical impact parameter and $L$ represents the radius  of unstable light ring.

 Eq.(\ref{cond-LR}) implies   two relations
 \begin{equation}
 L_i^2=f(L_i)b_i^2,\quad 2f(L_i)-L_if'(L_i)=0.
 \end{equation}
Here, four light rings and their critical impact parameters ($i=L,H,N,C$) for the EC are given by
\begin{eqnarray}
&&L_{i}(m,q,\mu), \label{LR} \\
&&b_{i}(m,q,\mu),\label{CI}
\end{eqnarray}
whose explicit forms are too complicated to write down here.
For a reference, we need to have the light ring and critical impact parameter for the RN as
\begin{equation}
L_{RN}(m,q)=\frac{3m}{2}\Big[1+\sqrt{1-\frac{8q^2}{9m^2}}\Big],\quad b_{RN}(m,q)=\frac{3m(1+\sqrt{1-8q^2/9m^2})}{\sqrt{2+\frac{3m^2(1+\sqrt{1-8q^2/9m^2})}{2q^2}}}.
\end{equation}
\begin{figure*}[t!]
   \centering
  \includegraphics[width=0.4\textwidth]{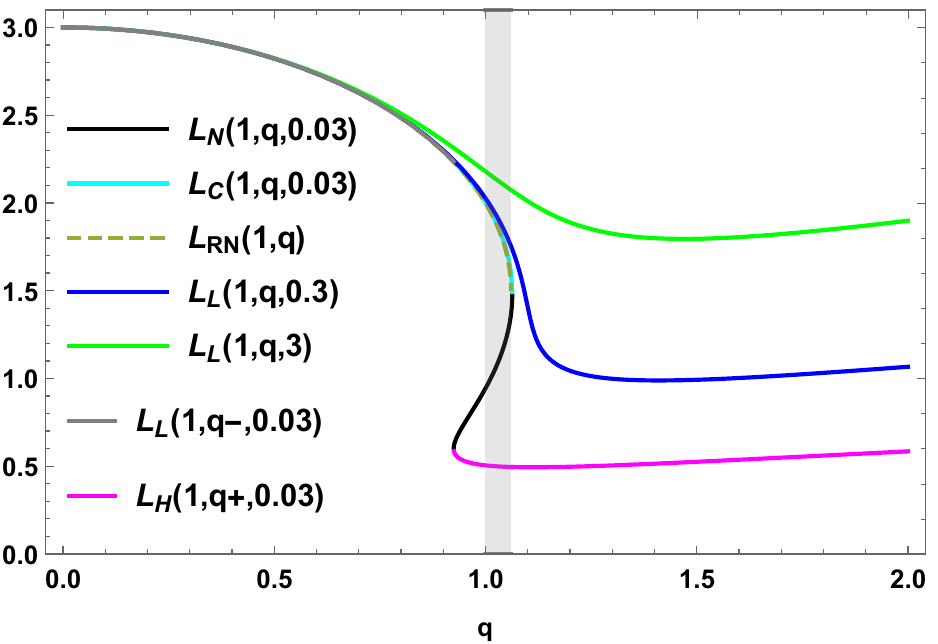}
  \hfill%
    \includegraphics[width=0.4\textwidth]{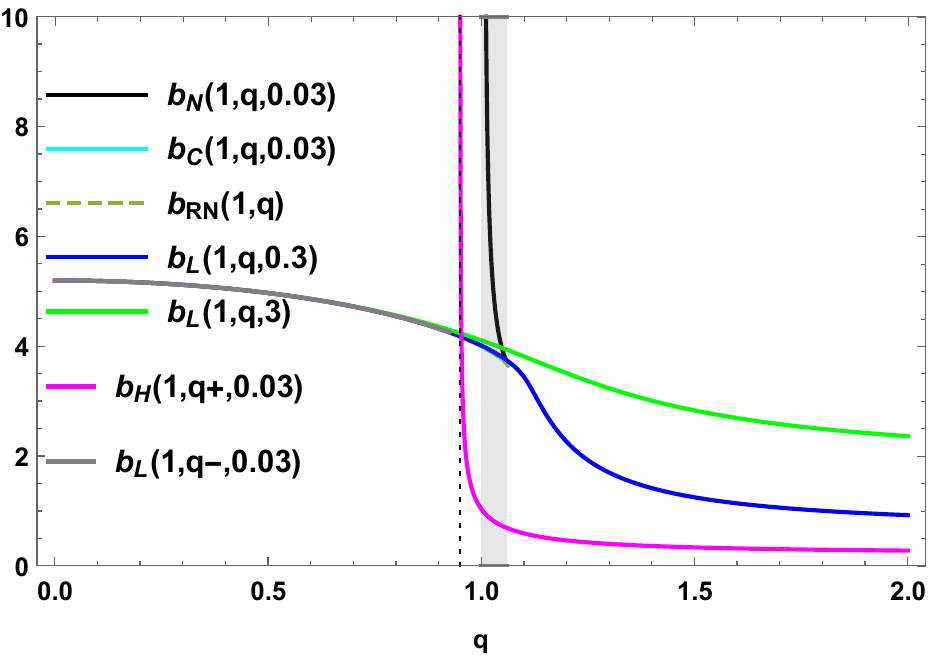}
\caption{ (Left)  Four light rings  $L_{i}(m=1,q,\mu=0.03)$   for $i$=L,H,N,C and  $L_L(1,q,0.3),~L_L(1,q,3),~L_{RN}(1,q)$ are  as functions of $q$.  The N, C, RN-branches are  extended to include their NS versions defined in the shaded column ($q\in[1,1.06]$).
(Right) Four critical impact parameters  $b_{i}(m=1,q,\mu=0.03)$   for $i$=L,H,N,C and  $b_L(1,q,0.3),~b_L(1,q,3),~b_{RN}(1,q)$ are  as functions of $q$. A dotted line represents $q=0.95$ which is a blow-up point for $b_{H}(1,q+,0.03)$. A shaded column ($q\in[1,1.06]$) includes  their N, C, RN-NS branches.}
\end{figure*}

(Left) Fig. 5 shows the light rings and (Right) Fig. 5 represents shadow radii.
We note that $L_{N}(1,q,0.03)$ and  $L_{C}(1,q,0.03)$ are  present as  connectors appearing  between $L_L(1,q-,0.03)$ and $L_{H}(1,q+,0.03)$, while  $L_{L}(1,q,0.3)$ and  $L_{L}(1,q,3)$ are single functions of $q$.
From analyzing  $L_{RN}(1,q)$, one finds  that  its $q$-range is extended  from $[0,1]$ to $[0,\sqrt{9/8}=1.06]$. This implies   that  $q\in[1,1.06]$ (shaded column) denotes the RN-NS branch.
Also, $L_N(1,q,0.03)$ and $L_C(1,q,0.03)$ are extended to enter their NS branches (N-NS,C-NS) into this column.
 Importantly, the dotted line in (Right) Fig. 5 indicates  the  blow-up  point for $b_{H}(1,q+,0.03)$ and a shaded column ($q\in[1,1.06]$) includes  their N, C, RN-NS branches.

\section{Test with EHT observation}
\begin{figure*}[t!]
   \centering
  \includegraphics[width=0.4\textwidth]{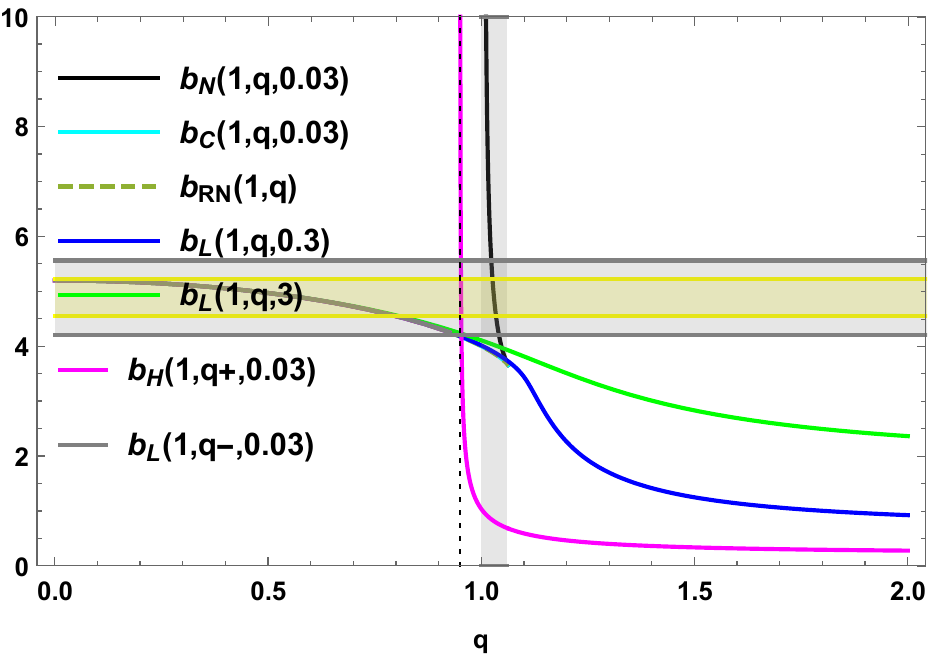}
 \hfill%
    \includegraphics[width=0.4\textwidth]{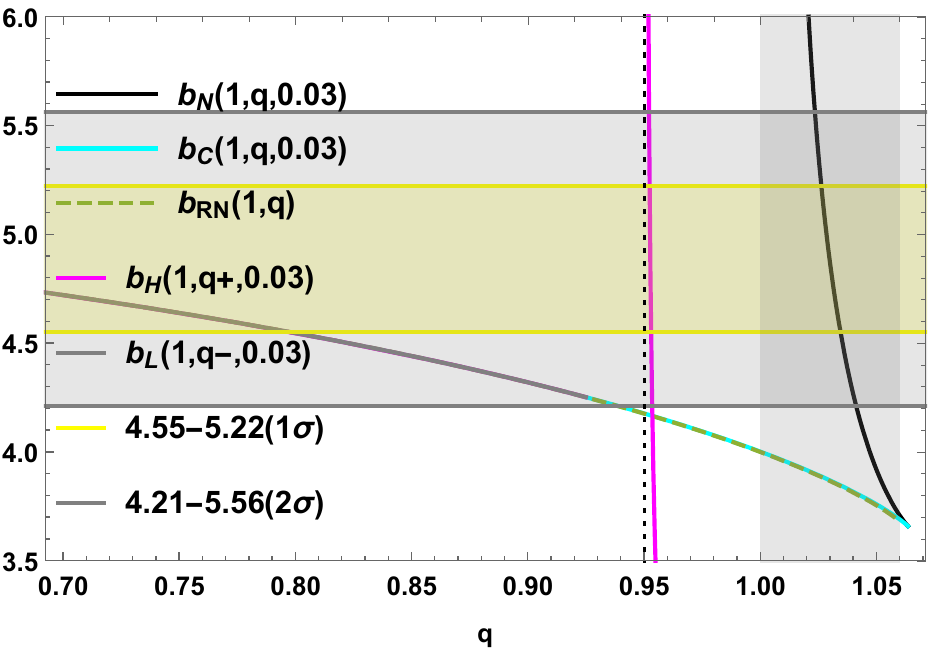}
\caption{(Left) Several  critical impact parameters  $b_i(m=1,q,\mu=0.03)$ for $i$=L,N,C,H, $b_{L}(1,q,0.3)$, $b_{L}(1,q,3)$, and $b_{RN}(1,q)$ as functions of $q\in[0,2]$.  There is one blow-up (dotted line) at $q=0.95$.
   Here, we introduce $1\sigma$ and $2\sigma$ ranges.
 (Right) Enlarged critical impact parameters $b_i(1,q,0.03)$ and $b_{RN}(1,q)$ are  functions of  $q\in[0.70,1.06]$. The dotted line at $q=0.95$  indicates a blow-up  for the hot branch. The shaded column includes the negative-NS,  cold-NS, and RN-NS  branches.   }
\end{figure*}
Consulting the EHT observation (Keck- and VLTI-based estimates for SgrA$*$~\cite{EventHorizonTelescope:2022wkp,EventHorizonTelescope:2022wok,EventHorizonTelescope:2022xqj}), the  $1\sigma$ constraint on the shadow radius $r_{\rm sh}=b_i$ indicates ~\cite{Vagnozzi:2022moj}
\begin{equation}
4.55\lesssim r_{\rm sh} \lesssim 5.22  \label{KV1}
\end{equation}
and the  $2\sigma$ constraint shows
\begin{equation}
4.21 \lesssim r_{\rm sh} \lesssim 5.56. \label{KV2}
\end{equation}
Let us see Fig. 6 for explicit pictures for testing with the EHT observation.
For the low branch of the EC, one has two constraints of the upper limits on its electric  charge $q$: $q\lesssim 0.798 (1\sigma)$ and 0.953 $(2\sigma)$ for $\mu=0.03$; $q\lesssim 0.799 (1\sigma)$ and 0.941 $(2\sigma)$ for $\mu=0.3$;
$q\lesssim 0.806 (1\sigma)$ and 0.961 $(2\sigma)$ for $\mu=3$.  Also, one has a narrow constraint of $0.925\lesssim q \lesssim 0.94 (2\sigma)$ for the cold branch. There is no constraint for the hot branch
because $b_{H}(1,q+,0.03)$ is a nearly vertical line.
 Similarly, the RN is constrained as $q\lesssim 0.798 (1\sigma)$ and  $0.939(2\sigma)$.
The EHT observation rules out the possibility of SgrA$^*$ being an extremal RN ($q_{eRN}=1$) and the RN-NS branch ($q\in[1,1.06]$: shaded column) is excluded from $q\lesssim 0.939(2\sigma)$~\cite{Vagnozzi:2022moj}.

From (Right) Fig. 6, one observes that two narrow ranges of  $1.026\lesssim q \lesssim 1.035 (1\sigma)$ and  $1.024\lesssim q \lesssim 1.041 (2\sigma)$ exist for the negative-NS branch.  We note that the cold-NS branch is ruled out by the $2\sigma$.
Also, it is worth  noting  that there is no $q>1$ range which constrains its magnetic  charge for $\mu=0.3,3$  because their critical impact parameters are monotonically decreasing functions of $q$ [see (Left) Fig. 6].

\section{Discussions}

The shadow radii of  various BH and NS found from modified gravity theories  were extensively  used to test the EHT results for SgrA$^*$ BH~\cite{EventHorizonTelescope:2022wkp,EventHorizonTelescope:2022wok,EventHorizonTelescope:2022xqj} and thus, to constrain their hair  parameters~\cite{Vagnozzi:2022moj}.
The MC was employed to investigating shadow radii~\cite{Allahyari:2019jqz,Vagnozzi:2022moj}.

In this work, we tested the EC for three coupling constants ($\mu=0.03,0.3,3$)  with the EHT observation for SgrA$^*$  by computing their shadow radii.
For this purpose, we performed thermodynamic analysis on the EC. For $\mu=0.03$, we clearly understand  what four branches of the low, the negative, the hot, and the cold represent  by considering their temperatures.
Computing their heat capacity leads to that  three branches of low, (negative), and RN   including Davies points  make phase transitions from negative (positive) to positive (negative) heat capacity.  It is found that the hot branch is always thermodynamically stable, whereas the cold  branch is always  thermodynamically unstable.    We showed  that  three branches of the low, the cold, and the hot  with RN one are globally stable, while the negative branch is globally unstable by computing the Helmholtz free energy including an extremal point as the ground state.
In this case, the shadow radius  for the low branch  is  the nearly same as that for the RN for $q<1$ case,   while  one finds  the $q>1$ negative-NS branch which could be constrained by the EHT observation.
A narrow constraint of $0.925 \lesssim q\lesssim 0.94(2\sigma)$ exists for the cold branch.
However, there is no  constraint on the  hot branch because $b_{H}(1,q+,0.03)$ is a nearly vertical line.  The extremal point of $q=1.0065$ is ruled out. 

For $\mu=0.3, 3$, there exist two single branches  without limitation on electric  charge $q$ whose shadow radii are monotonically decreasing functions of $q$. This indicates no extremal points. Their heat capacities are always negative, implying that they are always thermodynamically unstable. Also, their Helmholtz free energies are positive for $q\in[0,1.2]$, showing that they are globally unstable.   Their shadow radii are very similar to that of the RN for $q<1$, but there is no constraints for $q>1$.

\vspace{1cm}

{\bf Acknowledgments}
 \vspace{1cm}

 This work was supported by the National Research Foundation of Korea (NRF) grant
 funded by the Korea government(MSIT) (RS-2022-NR069013).

\end{document}